\documentclass{article}%
\usepackage{amssymb}
\usepackage{amsfonts}
\usepackage{amsmath}
\usepackage{graphicx}%
\begin{document}

\title{{\LARGE Negative capacitance and related }\\{\LARGE instabilities in theoretical models of the electric double layer and
membrane capacitors}}
\author{Michael B. Partenskii and Peter C. Jordan\\Department of Chemistry, MS-015\\Brandeis University\\PO Box 549110\\Waltham, MA 02454-9110, USA}
\maketitle

\begin{abstract}
Various models leading to predictions of negative capacitance, $C$, are
briefly reviewed. Their relation to the nature of electric control is
discussed. We reconfirm that the calculated double layer capacitance can be
negative under $\sigma$-control - an artificial construct that requires
uniform distribution of the electrode surface charge density, $\sigma
$\textbf{.\ }\ For instance, It is shown that the combined relaxation of the
ionic and electronic contributions can result in $C<0$\ even for the local
statistical ionic models with strictly positive diffuse layer capacitance. In
reality,\textbf{\ }however, only the total charge $q$ (or the average surface
charge density $\overline{\sigma}$) can be experimentally fixed in isolated
cell studies ($q$-control). For those $\sigma$ where $C$ becomes negative
under $\sigma$-control, the transition to $q$-control (i.e. relaxing the
lateral change density distribution, fixing its mean value to $\sigma$) leads
to instability of the uniform distribution and a transition to a non-uniform
phase. As an illustration, a "membrane capacitor" model is discussed. This
exactly solvable model, allowing for both uniform and inhomogeneous relaxation
of the electrical double layer, helps to demonstrate both the onset and some
important features of the instability. Possibilities for further development
are discussed briefly.

\end{abstract}

\section{Introduction}

\bigskip The admissible sign of the differential capacitance at charged
interfaces was questioned in early 1970s in relation to the so called
Cooper-Harrison catastrophe \cite{CooHar75}, an apparent prediction of $C<0$
for "dipolar capacitors". Soon thereafter, similar anomalies were predicted
for some ionic models \cite{Blu77,HenBluSmi79} and, somewhat later, for
microscopic "relaxing gap capacitor" (RGC) models which accounted for the
metal electron contributions in the interfacial capacitance
\ (see\ \cite{SchHen86,ParJor93,ParDorJor96,ParJor2001a} \ and references
therein) . It is generally accepted that $C$ must be strictly positive for
open "$\phi$-controlled" systems, where the electric cell is connected to a
source of controllable voltage \cite{LanLif70}. However, it was shown that
$C<0$ is possible as a stable state of an isolated RGC$\ $if electric control
is maintained by a macroscopically uniformly distributed surface charge
density $\sigma,$ so called $\sigma$-control (see
\cite{FelParVor86b,ParDorJor96,ParJor2001a} for review). Prior to this
discovery, an attempt had been made to prove, on general
statistical-mechanical grounds, that for an equilibrium $\sigma$-controlled
system $C$ must be strictly positive \cite{BluLebHen80}. Somewhat later
improved analysis \ \cite{ParFel89,NikAnaPap91,Nik92a,AttWeiPat92,ParJor93}
showed the model \cite{BluLebHen80} does not forbid negative $C$ values.
For\ both ionic and "relaxing gap capacitor" models, which together cover a
very wide range of interfacial phenomena, the equilibrium capacitance under
$\sigma$-control can be negative. This is true for both the "compact layer"
capacitance and the total double layer capacitance, including the\textbf{\ }%
"diffuse" layer contributions \cite{ParDorJor96,ParJor2001a}.

Here we show, that though acceptable for $\sigma$-control,  negative EDL
capacitance is not possible for a real isolated system, where uniformity of
$\sigma$ is not enforceable and only the total electrode charge can be fixed.
In other words, we show that in $\sigma$ domains where $C$ is negative under
$\sigma$-control, the system is unstable with respect to transition to an
inhomogeneous state with nonuniform lateral distributions of the electrode
charge density and of mobile electrolyte ions. To demonstrate this anomaly, we
discuss a transparent "membrane capacitor" model, which exhibits a
$C<0$\ domain presuming uniformity  while becoming unstable in this domain if
non-uniform surface charge distributions and membrane deformations are considered.

This result addresses questions raised recently
\cite{GonJimMes2004,BodHenPla2004} with respect to the meaning and physical
reality\ of $C<0$ for the isolated capacitor. Our analysis suggests that a
model demonstrating this anomaly under $\sigma$-control can be used to study
transition to an inhomogeneous interfacial state under $q$-control.

\section{Control of electrified interfaces - theory and experiment}

1. Experimental study of the electrical double layer (EDL) at electrochemical
interfaces is usually conducted under "potential" ($\phi$-) control, where
electrodes are connected to a voltage source. Changing the applied voltage in
increments $\Delta\phi$\ leads to corresponding changes of the electrode
charge, $\Delta q.$ Similarly, controlled modulation of the voltage,
$\phi=\phi(t)$, results in charge modulation measurable by impedance
techniques. Connection to a potentiostat, required to maintain $\phi$-control,
results in an open system, which is treated by grand canonical methods.

The differential capacitance (per unit area) is then defined by the derivative%
\begin{equation}
C_{\phi}=\partial_{\phi}\overline{\sigma} \label{C_Phi}%
\end{equation}
where the average surface charge density is $\overline{\sigma}=q/A$ and $A$ is
the the surface area of the electrode. \ The form of Eq. \ref{C_Phi} is a
typical response function
\begin{equation}
\varkappa_{F}=\partial_{F}X \label{resp}%
\end{equation}
where $F~$is the external parameter ("force") and $X~$the conjugate intensive variable.

2. The electrical properties of interfaces can equally well be studied by
controlling the electrode charge, $q$. In $q$-control the measured quantity is
the corresponding potential $\phi$. Charge can be regulated by connecting the
electrodes to a battery for brief periods of time $\Delta t$, measuring the
current $j$, with the charge increments found by $\Delta q=j\Delta t$. With
$q$ fixed, the resultant $\phi$ is measured in an isolated system, i.e. a
canonical ensemble.

The corresponding response function, analogous to Eqs. \ref{C_Phi} and
\ref{resp} is the inverse differential capacitance%

\begin{equation}
C_{q}^{-1}=\partial_{\overline{\sigma}}\phi. \label{C_q}%
\end{equation}
Obviously,\ $q$-control is\ a\ synonym\ for\ $\overline{\sigma}$%
-control:\ fixing\ the\ total\ charge\ $q~$%
is\ equivalent\ to\ fixing\ the\ average\ surface\ charge\ density\ $\overline
{\sigma}$. The thermodynamic potentials of the expanded ($\phi$-controlled )
system, $A_{\phi},$and the isolated ($q$-controlled) system, $A_{q},$ are
related by the Legendre transformation,
\begin{equation}
A_{\phi}(\overline{\sigma},\phi)=A_{q}(\overline{\sigma})-q\phi.
\label{Expand_Isolate}%
\end{equation}

3. Most calculations of the electric double layer (EDL) assume the electrode,
often described as a charged flat wall, has a uniform, fixed charge density.
\ This is effectively $\sigma$-control, which is generally not equivalent to
$q$- (or$~\overline{\sigma}$-)$\ $control. The terms are interchangeable only
if the equilibrium surface charge density is uniform on a scale exceeding
atomic dimensions, i.e. if $\,\sigma=\overline{\sigma}=const.$ A
counter-example is one where the optimized local charge density is non-uniform
in the electrode plane, $\sigma=\sigma(r_{s})$ $\neq\overline{\sigma}$
($r_{s}$\ is the radius vector in the electrode plane), reminiscent of "charge
density wave" states in electron plasma. In practice\textbf{\ }there is no way
to control the local charge density; in isolated systems only the total charge
can be constrained externally. $\ \sigma$-control is a purely theoretical
construct; its predictions must be tested to determine if a uniform surface
charge density and its corollary, a\textbf{\ }laterally-uniform ionic
distribution correspond to a real equilibrium state.

\section{Admissible sign of the differential capacitance}

\subsection{$\phi$-control: the open system}

General thermodynamic \cite{LanLif60} and statistical-mechanical \cite{McC71}
treatments of electrified interfaces show that, under $\phi$-control,
differential capacitance must be strictly positive. In our view attempts to
circumvent this restriction \cite{AttWeiPat92,Tor92,WeiTorPat93} have been
based on misinterpretations of the nature of potential control
\cite{ParJor93,ParJor2001a}. Near a critical voltage $\phi_{cr}$, defined by
$C^{-1}(\phi_{cr})=0$, the system becomes unstable. The transition to a new
state is accompanied by charge flow from the potentiostat to the electrodes, a
sort of electrical "breakdown" (see
\cite{FelParVor86a,FelParVor86b,ParDorJor96} for more details)\textbf{, }which
would be a unique path to phase transformation assuming lateral uniformity.
However, as discussed below, the transition can also involve formation of a
laterally non-uniform phase accompanied by nonuniform redistribution of the
electrode charge density\textbf{\ }$\sigma(\rho)$\textbf{. }

As the requirement that $C_{\phi}>0$\ \ is now generally accepted, we turn to
treating isolated systems. \ The sense of the upcoming discussion is already
implicit, once having recognized\textbf{\ }that the\textbf{\ }first of these
instabilities is forbidden by the very definition of "$q$-control" since, once
$q$\ is fixed, electric contact with the potentiostat must be interrupted. Can
$C$\ be negative under this constraint? \ As the admissible sign of $C$\ in
isolated systems has\ almost always been analyzed in $\sigma$-control terms,
we first consider this case and postpone discussion of the more general $q$-control.

\subsection{$\sigma$-control in the isolated system}

\subsubsection{Primitive models of electrolytes}

Interest in the admissible sign of $C$ in the theory of the diffusive layer
was stimulated by work of Blum, Lebowitz and Henderson \cite{BluLebHen80}.
They tried to provide a rigorous restriction on the sign of $C$ for "primitive
ionic models:" charged hard ions in a uniform dielectric medium between two
rigid, uniformly charged walls.\textbf{\ }The corresponding Hamiltonian is
quite generally \cite{ParFel89}
\begin{equation}
H(\sigma,\{\mathbf{R}\})=\frac{\sigma^{2}d}{2\varepsilon\varepsilon_{0}%
}-\sigma f(\{\mathbf{R}\})+H^{^{\prime}}(\{\mathbf{R}\}) \label{Hamilt-Isol}%
\end{equation}
where $\{\mathbf{R}\}$ \ refers to a particular configuration of the system
(charge coordinates, dipolar orientations, etc.). The first term describes
direct interaction between the charged walls, with $d$ the inter-wall
distance. $\sigma f(\{\mathbf{R}\})~$accounts for interaction between the
electrolyte and the electrode field (the physical significance of $f$ will be
clear shortly) and $H^{\prime}$ is a $\sigma$ independent interaction energy.
The potential drop between the charged plates is%
\begin{equation}
\phi(\sigma)=\frac{\sigma d}{\varepsilon\varepsilon_{0}}~+\ <f> \label{phi}%
\end{equation}
where%
\[
<(...)>=\frac{\int e^{-\beta H(\sigma,\{\mathbf{R}\})}(...)d\Omega}{\int
e^{-\beta H(\sigma,\{\mathbf{R}\})}d\Omega}%
\]
is a canonical average with integration over the system's configurational
space, $\Omega$, $\ $and $\beta=1/kT.$ Eq. \ref{phi} reveals the meaning of
$\ f$: $\ <f>$\ is the potential drop induced in the electrolyte by the field
of the charged plates. It arises from redistribution of free (ionic) charges
shielding the applied field, and from repositioning of the bound charges ( the
reorientation of molecular dipoles).

\ For Hamiltonians of the type of Eq. \ref{Hamilt-Isol} the capacitance
satisfies the general condition \cite{ParFel89}
\begin{equation}
C^{-1}=\frac{d}{\varepsilon\varepsilon_{0}}-\frac{A}{kT}(<f^{~2}>-<f>^{2}),
\label{C_Isol}%
\end{equation}
leading to a simple and self-evident result:\ ~%
\begin{equation}
C^{-1}\leq\frac{d}{\varepsilon\varepsilon_{0}}. \label{C<=}%
\end{equation}

Eq. \ref{C<=} simply means that redistribution of free charges and molecular
polarization induced by the electric field in the electrolyte reduce the
potential drop between the electrodes and increase the capacitance $C$, a
condition that places no constraints on the sign of $C$ under $\sigma
$-control\textit{. \ }One should note here that if the distance $d$
$>>\lambda_{D}$, the characteristic Debye length in the electrolyte, then the
total inverse capacitance of the cell splits into two independent double layer
contributions belonging to two "electrodes:"%

\[
C^{-1}=C_{1}^{-1}+C_{2}^{-1}.
\]
If the sign of the total cell capacitance\ $C$ is unrestricted under $\sigma
$-control, this is even more true for the individual double layer
contributions, $C_{1}$\ and $C_{2}$.

\textbf{\ }Eq. \ref{C_Isol} (Eq. 24 of \cite{ParFel89}) was derived in a study
of a "dipolar capacitor" ("DC"), a lattice of point dipoles embedded between
the plates of a parallel-plate capacitor, a model often used for analyzing a
compact layer at metal-solvent interfaces. For the DC%

\[
f~=f_{_{DC}}=-\frac{1}{\varepsilon_{0}}P_{z}=\frac{1}{A\varepsilon_{0}}%
\sum_{i}p_{i,z}%
\]
is the potential drop corresponding to an arbitrary configuration of the
molecular dipoles with $P_{z}\ $the average surface density of the dipole
moment in the lattice and $p_{i,z}$ the projection of the individual dipole
moment normal to the surface of the lattice.

If interaction between the charged walls (the first contribution to the
Hamiltonian, Eq. \ref{Hamilt-Isol}) were properly included in the Hamiltonian
of the primitive ionic model \cite{BluLebHen80}, it \ would also lead to Eq.
\ref{C_Isol} (see p. 68 of\ \cite{ParFel89}) with%

\[
f=f_{ion}=-\frac{1}{\varepsilon_{0}}\sum q_{i}z_{i}%
\]
where $q_{i\text{ \ }}$is the charge of $i$-th ion and $z_{i}$ its distance
from the charged wall positioned at $z=0$. \ Eq. \ref{C_Isol} \ and its
analogs have been repeatedly derived and discussed
\cite{NikAnaPap91,AttWeiPat92,ParJor93}; they hold for any model in which the
electrodes are treated as hard charged walls with distinctly separate regions
occupied by electrode and electrolyte. With these restrictions\textbf{\ }the
interaction between the electrode and the electrolyte can quite
generally\textbf{\ }be described by a contribution $\sim\sigma\sum q_{i}z_{i}$
where the summation includes both the mobile ionic and the molecular multipole
charges \cite{ParJor93}. Such constraints clearly exclude "polarizable"
models, \ those explicitly treating molecular electronic polarizability,
electron density penetration into regions occupied by electrolyte, etc.

\subsubsection{Relaxing gap capacitors (RGC)}

Immobility of the charged "plates" in primitive models does not account for
another important phenomenon, possible displacement of the "electronic plate"
of interfacial capacitors and of the equilibrium positions of the electrolyte
species in contact with the electrode, in response to charging
\cite{SchHen86,ParJor93,ParDorJor96,ParJor2001a}. These effects
are\ effectively\textbf{\ }illustrated by the "relaxing gap capacitor"
\ metaphor, which emphasizes the dependence of the effective gap $d~$on
charging. The potential drop can be quite generally represented as%

\begin{equation}
\phi=\frac{1}{\varepsilon\varepsilon_{0}}\sigma d(\sigma) \label{poten_RGC}%
\end{equation}
where $d(\sigma)$ is the effective separation between the "plates" of the
capacitor associated with the "centers of mass" of two microscopic charge
distributions (see \cite{SchHen86,FelParVor86b,ParDorJor96} for review). \ We
assume a uniform dielectric background $\varepsilon$ between the plates, in
the volume occupied by free charges. For models using a non-uniform
background, typical of unified models used to account for both the "inner" (or
Helmholtz) and the diffuse layers \cite{BocRedGam2000}, the expression is more
complex and must also account for the distributions of the bound
(polarization) charges. However, these details are not essential; they are
omitted here.

For the electrode-electrolyte interface the effective separation is%

\begin{equation}
d(\sigma)=z_{i}(\sigma)-z_{i,e}(\sigma)
\end{equation}
where
\[
z_{i,e}=\frac{\int\rho_{i,e}^{\sigma}zdz}{\sigma}%
\]
with $z$ the coordinate normal to the electrode surface. The inverse
differential capacitance for the RGC is%
\begin{equation}
(\frac{1}{\varepsilon\varepsilon_{0}}C)^{-1}=(\sigma d(\sigma))_{\sigma
}^{\prime}=d(\sigma)+\sigma d(\sigma)_{\sigma}^{\prime}. \label{cap_relaxed}%
\end{equation}
Dependence of $d\ $on $\sigma$, a general feature of practically all double
layer models, implies that $C$ is dependent on $\sigma~($or on the applied
voltage). The "plate" displacement that contributes to the variation of $d$
\ not only reflects a shift of charge density profiles, but is more a
consequence of shape variation \cite{ParJor2001a}. Quite typically, there is
always a $\sigma$ range in which charging decreases the\ effective gap.
Elastic compression of the lipid membrane by electric stress
\cite{Cro73,ParDorJor98b,ParJor2000b} and response of the Gouy-Chapman-Stern
(GCS) diffuse layer to charging \cite{BocRedGam2000} are two representative examples.

In a range of $\sigma$ where the effective gap contracts with charging,
$d^{^{\prime}}(\sigma)<0$ and $C_{\sigma}$ can be negative if%

\begin{equation}
\Delta\sigma\cdot d-\sigma\cdot\Delta d<0. \label{C<0 _cond}%
\end{equation}
This inequality means that a potential increase due to a change of $\sigma$ is
overwhelmed by its decrease due to gap contraction. A number of
electromechanical and microscopic models
\cite{FelParVor86b,ParJor93,ParDorJor96,ParJor2001a} show that negative
capacitance (NC)\ under $\sigma$-control is compatible with system stability.
For instance, for every\textit{\ fixed }$\sigma$ (including the domain where
$C<0$) the equilibrium gap of the elastic capacitor is defined by a stable
balance between elastic and electrostatic forces. Similarly, density
functional minimization led to a NC at metal-electrolyte interfaces due to
relaxation of the "electronic plate" of the capacitor
\cite{FelParVor86a,FelParVor86b,KimKorPar89}.

Previous work \cite{FelParVor86b,ParDorJor96} showed that accounting for
"electronic plate" relaxation in combination with traditional GCS and similar
models typically leads to negative $C$ domains even though the GCS model
itself (as with any other "local" statistical model where ionic concentrations
are local functions of the potential) does not lead to such anomalies
\cite{FelParVor86b}, a point further demonstrated in the Appendix. Thus, while
we agree that finding $C<0$ \textbf{\ }under $\sigma$-control must be both
common and important for ionic models of electrolytes \cite{GonJimMes2004}, it
is not a\ necessary condition for the appearance of this anomaly. If other
relaxation mechanisms are taken in account a NC\textbf{\ }domain may arise
even if the ionic contribution is positive.

\subsubsection{Possible capacitance anomalies for local $\sigma$-controlled
statistical models with interfacial relaxation}

Consider a conventional two-layer model of the double layer, with a Helmholtz
layer accounting for the finite ion-electrode distance of closest approach,
$a$, and the diffuse layer accounting for the electrolyte's ionic charge
distribution \cite{BocRedGam2000,BruGouPin94}. Its inverse capacitance is
\begin{equation}
C^{-1}(\sigma)=C_{H}^{-1}(\sigma)+C_{D}^{-1}(\sigma) \label{DLC}%
\end{equation}
Here
\begin{align}
C_{H}^{-1}(\sigma)  &  =\partial_{\sigma}\phi_{H}\label{DLC_comp}\\
C_{D}^{-1}(\sigma)  &  =\partial_{\sigma}\phi_{D}\nonumber
\end{align}
with $\phi_{H}$ and $\phi_{D}$ the potential drops in the Helmholtz and
diffuse layers respectively. We now treat "local statistical" diffuse layer models.

These simple models describe the ionic density $\rho_{i}$ at a distance $z$
from the electrode as a function of the local potential $\varphi(z)$:%
\begin{equation}
\rho_{i}=\rho_{i}[\varphi(z)] \label{LocModDens}%
\end{equation}
The classic example is the Poisson-Boltzman-Gouy-Chapman (PBGC) model of
symmetric electrolytes%

\begin{equation}
\rho_{i}=qn_{0}~\cos h[-\beta q\varphi(z)] \label{PBD}%
\end{equation}
with $n_{0\text{\ }}$the bulk concentration of cations or anions. For the
slightly more complex model treating "ionic saturation," finite ion size
driven entropic restriction on local ionic concentration, which can be
especially significant in solid electrolytes, the ionic density is%

\begin{equation}
\rho_{i}(z)=qn_{0}~[\frac{\exp[-\beta\varphi]}{1-\Theta_{1}+\Theta_{1}%
\exp[-\beta\varphi]}-\frac{\exp[\beta\varphi]}{1-\Theta_{2}+\Theta_{2}%
\exp[-\beta\varphi]} \label{PBD_satur}%
\end{equation}
where $\Theta_{1,2}=n_{0}/N_{1,2}~$and $N_{1,2}$ are maximal possible
concentrations for the cations and anions respectively. Concentration
limitations reflect finite ionic size and, for solid electrolytes, the
limitations on the number of possible ionic defect sites in the crystal lattice.

The one-dimensional contact model is illustrated in Fig. 1.\newline%

%TCIMACRO{\FRAME{ftbpF}{5.0678in}{2.5322in}{0pt}{}{}{fig1.eps}%
%width 5.0678in;  height 2.5322in;  depth 0pt;  original-width 5.0125in;
%original-height 2.4915in;  cropleft "0";  croptop "1";  cropright "1";
%filename 'fig1.eps';file-properties "XNPEU";}}}%
%BeginExpansion
\begin{figure}
[ptb]
\begin{center}
\includegraphics[
height=2.5322in,
width=5.0678in
]%
{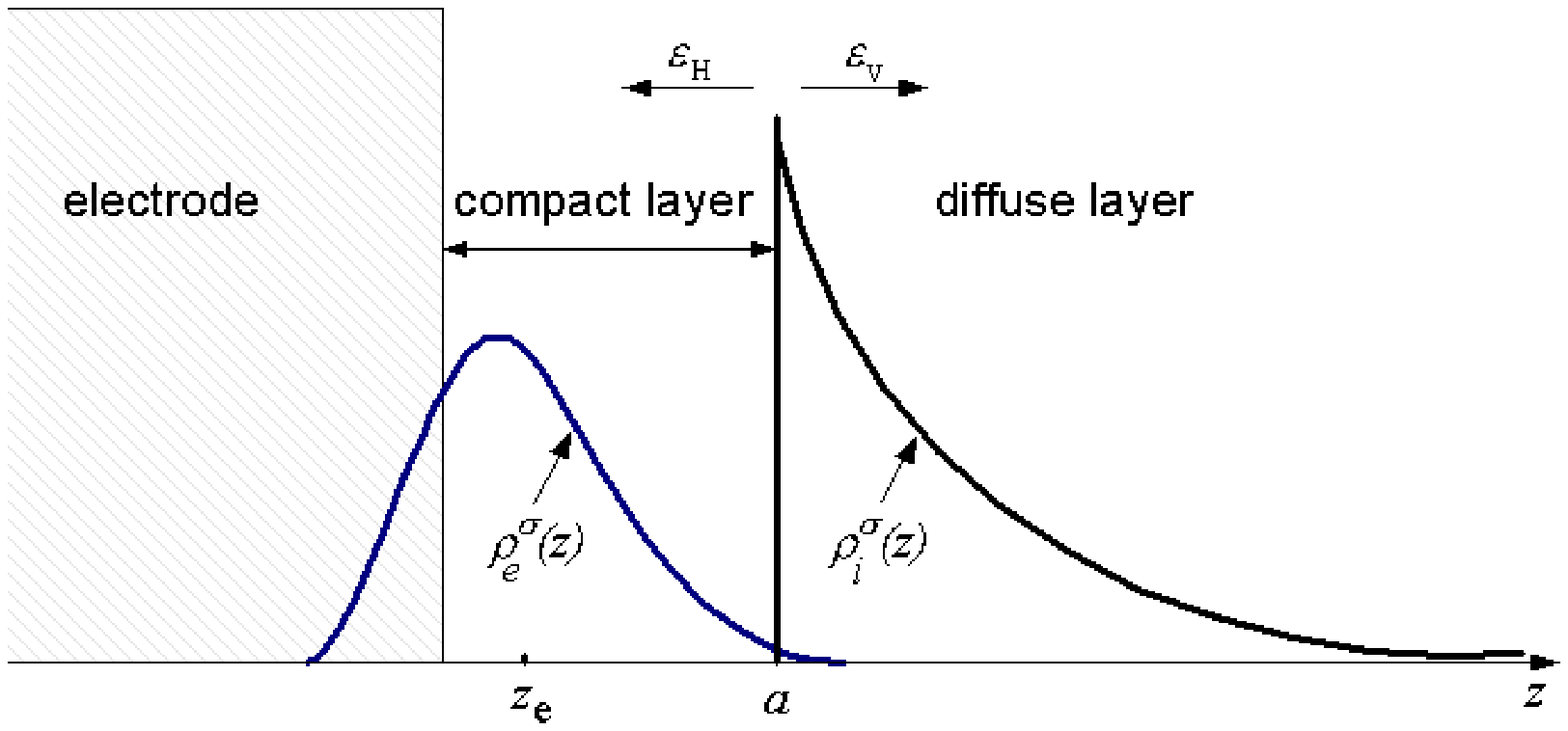}%
\end{center}
\end{figure}
%EndExpansion

\begin{quote}
\textbf{Fig. 1 }Schematic representation of the ionic, $\rho_{i}^{\sigma}%
,$\ and the electronic, $\rho_{e}^{\sigma}$\ , charge distributions induced by
the electrode surface charge density $\sigma.$

\end{quote}

We use a two-layer dielectric model of the interface%
\begin{equation}
\varepsilon(z)=\left\{
\begin{array}
[c]{c}%
\varepsilon_{H},\ \ \ z\leq a\ \\
\varepsilon_{v},\ \ z>a
\end{array}
\right.  \label{DLEps}%
\end{equation}
where $\varepsilon_{H}\ \ $and $\varepsilon_{v}\ $are dielectric constants of
the Helmholtz layer and bulk electrolyte respectively. Ignoring the
penetration of the electron density into the "diffuse" layer the potential
drop in the Helmholtz layer is
\begin{equation}
\phi_{H}=\frac{\sigma}{\varepsilon_{H}\varepsilon_{0}}d_{H}(\sigma)
\label{Phi_H}%
\end{equation}
where%
\begin{equation}
d_{H}(\sigma)=a-z_{e}(\sigma) \label{d_H}%
\end{equation}
is the effective gap of the Helmholtz layer. As a result, in direct analogy
with Eq. \ref{cap_relaxed},%

\begin{equation}
C_{H}^{-1}(\sigma)=\frac{1}{\varepsilon_{H}\varepsilon_{0}}[a-z_{e}%
(\sigma)-\sigma~\partial_{\sigma}\ z_{e}] \label{C_H- general}%
\end{equation}

The significance of Eq. \ref{Phi_H} is that $d_{H}(\sigma)$ depends on
$\sigma$ via displacement of the "electronic plate" $z_{e}$ of the equivalent
interfacial capacitor during the charging\textbf{\ }process. This effectively
turns $C_{H}$ into a "relaxing gap capacitor"
\cite{SchHen86,FelParVor86b,ParDorJor96}. The $\sigma$-dependence of the
position of the equivalent electronic plate has been studied using a density
functional approach for\textbf{\ }various models of the ionic charge
distribution near the electrode, from the Raleigh picture \cite{Ral76} where
the ionic charge density is localized in one or two monolayers nearest the
electrode\textbf{, }to the "external field" limit, where countercharges are
located far from the electrode \cite{FelParVor86a,FelParVor86b,SchHen86} ).
The diffuse layer distribution is generally bounded by these limits;
$z_{e}(\sigma)$ is always well approximated by the cubic polynomial \
\begin{equation}
z_{e}(\sigma)=z_{e}(0)+s\ \sigma+p\sigma^{2}+r\sigma^{3} \label{z_e(sig)}%
\end{equation}

\begin{equation}
C_{H}^{-1}=\frac{1}{\varepsilon_{H}\varepsilon_{0}}(a-z_{e}(0)-2s\ \sigma
+3p\sigma^{2}) \label{C_F-final}%
\end{equation}

The diffuse layer ionic distribution is related to $\sigma$ via the strict
"sum rule"%

\begin{equation}
\sigma^{2}=-2\varepsilon_{v}\varepsilon_{0}\int_{0}^{\phi_{d}}\rho
(\varphi)d\varphi\label{Sig_ro_diffuse}%
\end{equation}
which leads to a general expression for the capacitance%

\[
C_{D}^{-1}=-\frac{1}{\varepsilon_{v}\varepsilon_{0}}\frac{\sigma}{\rho
_{i}(\phi_{D})}%
\]
where $\rho_{i}(\phi_{D})=\rho_{i}(a)\ $(see Eq. \ref{LocModDens} ) is the
ionic charge density at $\ z=a$, the point where the local potential $\varphi$
equals $\phi_{D}$ (we choose $\varphi(\infty)=0$). Since $\sigma$ and
$\rho_{i}(a)$ are of opposite sign, this is simply%

\begin{equation}
C_{D}^{-1}=\frac{1}{\varepsilon_{v}\varepsilon_{0}}\left\vert \frac{\sigma
}{\rho_{i}(\phi_{D})}\right\vert , \label{Cd_exact_Mod}%
\end{equation}
which shows that $C_{D}$ is always positive and finite for any finite $\sigma
$. Thus local statistical models don't satisfy the criteria suggested by
\cite{GonJimMes2004}. We now show that charging induced relaxation of the
"electron plate" (see Eqs. \ref{z_e(sig)} and \ref{C_F-final}) can lead to a
negative total capacitance even if the diffusive contribution is strictly
positive. In the spirit of $\sigma$-control we express $\phi_{D}$ and $C_{D}$
solely in terms of $\sigma$. Corresponding relations were discussed for
various local ionic models in \ \cite{ParKha89, KimParSol89}. Thus for the
model of Eq. \ref{PBD_satur} the result is
\begin{equation}
(1-\Theta_{1}+\Theta_{1}u)(1-\Theta_{1}+\Theta_{1}u^{-1})^{\alpha}=G(\sigma)
\label{Pot_Char_Eq}%
\end{equation}
where
\[
u=\exp\left(  -\frac{q\phi_{D})}{kT}\right)  ,\ \ G(\sigma)=\exp\left(
\frac{1}{2\varepsilon\varepsilon_{0}N_{1}kT}~\sigma^{2}\right)  ,~\alpha
=\frac{\Theta_{1}}{\Theta_{2}}=\frac{N_{2}}{N_{1}}%
\]

For the case $N_{1}=N_{2}=N$, corresponding to a solvent with cations and
anions of equal solvation shell radii or solid electrolytes with Shottky
defects, this can be solved \ analytically. The capacitance is
\begin{equation}
C_{D}~=\varepsilon\varepsilon_{0}n_{0}\frac{q}{|\sigma|G(\sigma)}\sqrt
{S^{2}+4S(1-\Theta_{0})} \label{C_D_Sigma}%
\end{equation}
where $S=(G-1)/\Theta_{0},~\Theta_{0}=n_{0}/N.\ $For $N_{1}\neq N_{2}$, Eq.
\ref{Pot_Char_Eq}$\ $must be $\ $solved numerically. \ We limit consideration
to comparatively low ionic concentrations where restrictions on ionic packing
can be neglected and the ionic distribution is described by the PBGC model,
Eq. \ref{PBD}. Then Eq. \ref{C_D_Sigma} leads to the familiar expression%

\begin{equation}
C_{D}^{PBGC}=\alpha~(\beta+\sigma^{2})^{1/2}=\frac{\varepsilon\varepsilon_{0}%
}{L_{D}}[1+(\sigma/\sigma_{0})^{2}]^{1/2} \label{CPB_sig}%
\end{equation}
with%
\begin{equation}
\ L_{D}=\sqrt{\frac{\varepsilon\varepsilon_{0}kT\ }{2q^{2}n_{0}}%
=}\ 2.82\ 10^{-12}\sqrt{\frac{\varepsilon T\ }{c_{0}}}[m],\ \ \sigma_{0}%
=\sqrt{8n_{0}\varepsilon\varepsilon_{0}kT}=7.67~10^{-4}\sqrt{c_{0}\varepsilon
T\ \ }\ \ [C/m^{2}] \label{LD}%
\end{equation}
where $c_{0}$ is the molarity of the solvent. In the $\sigma=0$ limit the
capacitance is naturally determined by the electrolyte's Debye length.

To explore the analogy with the relaxing gap capacitor, it is instructive to
represent $(C_{D}^{PBGC})^{-1}$as $\symbol{126}\partial_{\sigma}(\sigma
l(\sigma)~)$\ and determine the $\sigma-$dependence of the effective gap $l.$
$\ $Solving
\[
\sigma l(\sigma)\ ^{\prime}+l(\sigma)=\frac{L_{D}}{[1+(\sigma/\sigma_{0}%
)^{2}]^{1/2}}%
\]
yields%

\begin{equation}
l(\widetilde{\sigma})=\frac{L_{D}}{\Sigma}\ Log\left\vert \Sigma
+\sqrt{1+\Sigma^{2}}\right\vert \label{Diff_Gap}%
\end{equation}
where $\Sigma=\sigma/\sigma_{0}.$ The dimensionless gap relaxation $\alpha=l$
$/L_{D\text{ }}$ (line 1) and the corresponding dimensionless potential
$\Sigma\alpha(\Sigma)$ (line 2) are shown in Fig. 2 and compared with similar
features of the elastic capacitor ( lines 3 and 4).

\textbf{\
%TCIMACRO{\FRAME{ftbpF}{3.0441in}{3.9306in}{0pt}{}{}{fig2.eps}%
%width 3.0441in;  height 3.9306in;  depth 0pt;  original-width 3in;
%original-height 3.8821in;  cropleft "0";  croptop "1";  cropright "1";
%filename 'fig2.eps';file-properties "XNPEU";}}}%
%BeginExpansion
\begin{figure}
[ptb]
\begin{center}
\includegraphics[
height=3.9306in,
width=3.0441in
]%
{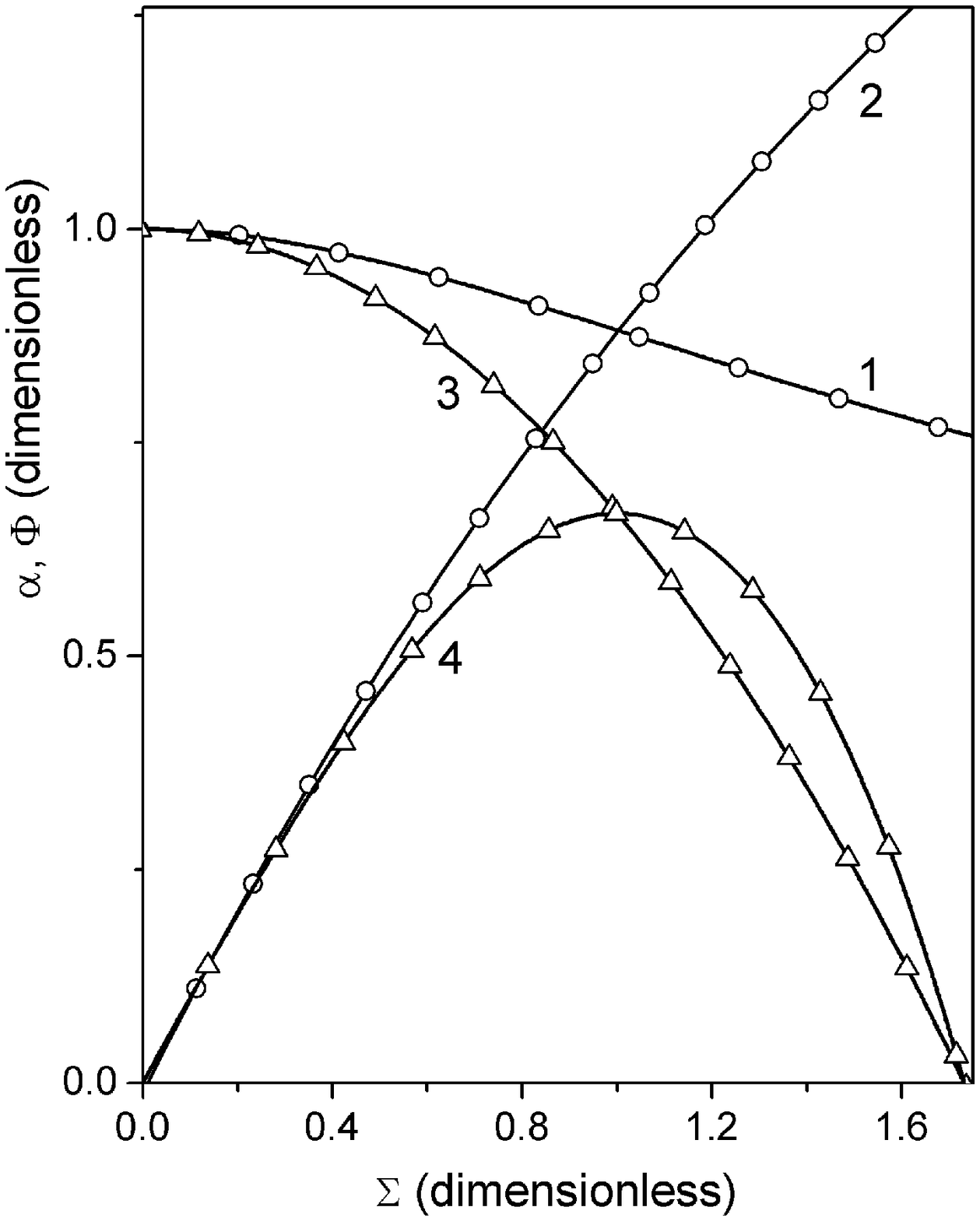}%
\end{center}
\end{figure}
%EndExpansion
}

\begin{quote}
\textbf{Fig. 2.} Charge dependence of the dimensionless effective gap
$\alpha(\Sigma)$ (curves 1 and 3) and the potential $\Phi=\Sigma\alpha
(\Sigma)\ $(curves 2 and 4) for the PBGC diffuse layer (curves 1 and 2) and
the membrane capacitor (curves 3 and 4); $\Sigma$\ is the dimensionless charge
density. For the PBGC diffuse layer $\alpha(\Sigma)\ =l(\Sigma)/L_{D\text{ }}%
$(see Eq. \ref{Diff_Gap} and the definition below) and for the membrane
capacitor $\alpha(\Sigma)\ =1-\frac{1}{3}\Sigma^{2}$\ (compare with Eq.
\ref{thinning-charge} below)\bigskip
\end{quote}

This picture clearly demonstrates how the relaxing gap capacitor is related to
the GC diffuse layer. In both models charging contracts the effective gap.
However, the rate of contraction in the PBGC model is insufficient for the
formation of a peak in $\phi(\Sigma)~$(curve 2), \ observed in the elastic
capacitor (curve 4) and associated with the $C<0$\textbf{\ }domain. This
follows directly from a general property of all local models, that $C$ is
positive at all surface charge densities, $\sigma$.

By considering the solid electrolyte, AgCl, for which the compact layer
electronic properties have been studied in \cite{KimParSol89, ParKha89}, we
show that a compact layer can lead to a negative total (diffuse + compact)
layer capacitance. The temperature dependence of the parameters $\varepsilon$
and $c_{0}$ are given by \cite{Mul67}.~At $400^{0}C$, $c_{0}\sim\ 43\ mM$ and
$\varepsilon\sim3$ so that $L_{D}\sim2.7$ \AA \ and $\sigma_{0}\sim
0.55\,\ \mu$C/cm$^{2}.$ Combining Eqs. \ref{C_F-final} and \ref{CPB_sig} we
find:$\bigskip$%
\begin{equation}
\varepsilon_{0}C^{-1}=\frac{1}{\varepsilon_{v}}\frac{L_{D}}{[1+(\sigma
/\sigma_{0})^{2}]^{1/2}}+\frac{1}{\varepsilon_{H}}(a-x_{e}(0)-2s\ \sigma
-3p\sigma^{2}-4rs^{3}) \label{C_relax_total}%
\end{equation}
According to the commonly held view, the contribution of the compact layer
effectively increases the effective gap, thus increasing the total inverse
capacitance. However, the appearance of a domain where $C<0$\ implies that
$C^{-1}$can be reduced sufficiently to become negative. Therefore ,\ one
expects that if local model diffuse layer contribution is strictly positive,
this should be even more pronounced in presence of the compact layer. The
following discussion demonstrate this to be wrong. Results for the double
layer capacitance are presented in Fig. 3.

The traditional model (line 1) includes the diffuse layer, Eq. \ref{CPB_sig},
in series with the fixed compact layer contribution calculated in the "perfect
conductor" model. This ignores diffuseness of the electronic distribution and
its relaxation in the compact layer, in effect assuming%
\[
C_{H}^{-1}=\frac{\varepsilon_{0}\varepsilon_{H}}{a}%
\]
Naturally this capacitance is positive at all $\sigma$. Accounting for
electronic relaxation in the compact layer dramatically affects behavior. The
parameters $x_{e}(o),s,~p~$and $r$ (Eqs \ref{z_e(sig)} and \ref{C_F-final})
were calculated by \cite{KimParSol89, ParKha89} for Au/AgX contacts.

\begin{quote}%
%TCIMACRO{\FRAME{ftbpF}{3.0441in}{3.9306in}{0pt}{}{}{fig3.eps}%
%width 3.0441in;  height 3.9306in;  depth 0pt;  original-width 3in;
%original-height 3.8821in;  cropleft "0";  croptop "1";  cropright "1";
%filename 'fig3.eps';file-properties "XNPEU";}}}%
%BeginExpansion
\begin{figure}
[ptb]
\begin{center}
\includegraphics[
height=3.9306in,
width=3.0441in
]%
{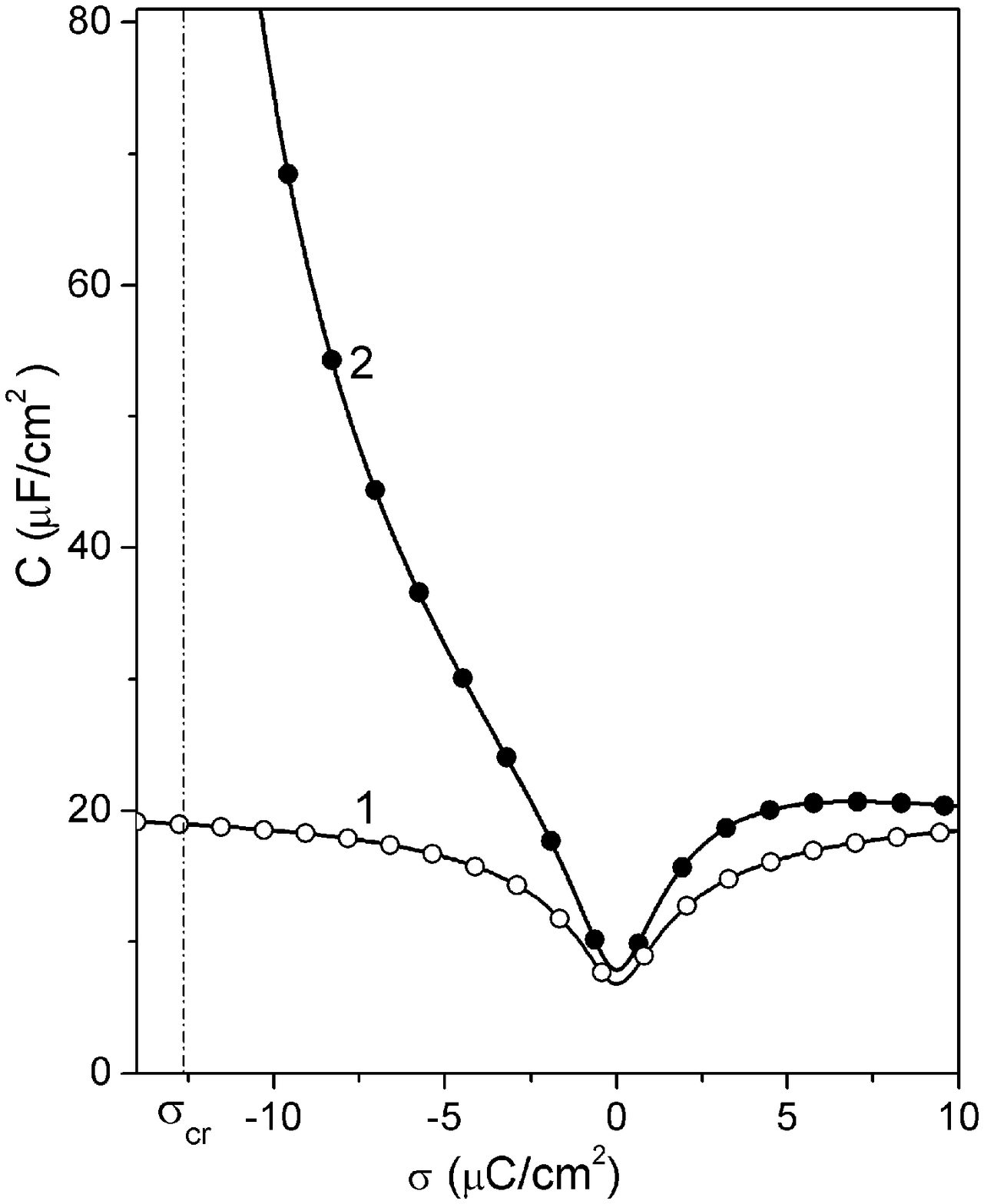}%
\end{center}
\end{figure}
%EndExpansion
\textbf{Fig. 3.} Double layer capacitance $C$\ for two models of the
interface:1 - traditional PBGC-Helmholtz model (the position of the "effective
electronic plate" is fixed at $z_{e}=0$); 2 - modified PBGC-Helmholtz model,
which also accounts for relaxation of the effective "electronic
plate."\textbf{\ \ }
\end{quote}

\begin{center}

\end{center}

The corresponding $C$ is shown for AgCl (line 2). Electronic relaxation leads
to a pronounced capacitance asymmetry and steep growth in the cathodic range
of charges, both effects observed experimentally \cite{Ral76,RemChe84}. It
yields a vertical $C$ asymptote separating $C>0$ ($\sigma>\sigma_{cr}) $ and
$C<0$ (not shown) domains. This result is general, and unaffected by variation
of $\varepsilon_{H}$ (typically from 2 to 5), $\varepsilon_{v}$ (from 3 to 10)
or electronic parameters reflecting different approximations to the electronic
density functional \cite{KimParSol89,ParKha89,FelParVor86a}. The appearance of
a negative capacitance domain is similar to phenomena arising from the compact
layer at metal-solvent interfaces \cite{KimKorPar89} \ Thus, $C<0$ may occur
under $\sigma-$control even for local diffuse layer models if the metal
electrode modeled realistically. Put differently, a $C<0$ domain may arise
even for local ionic models with strictly positive diffuse layer
contributions, if additional electronic (and possibly other) mechanisms for
interfacial relaxation are considered.

\bigskip

\subsection{$\mathbf{q}$\textbf{-control in the isolated system}}

As discussed, $\sigma$-control is an artificial construct. Nevertheless,
charges are normally distributed uniformly in the plane of the electrode and
the charge $q$ naturally yields a uniform charge density $\sigma$, in which
case $\sigma$- and $q$-control are identical. However, it is possible that
under special conditions a non-uniform distribution of charge in the electrode
plane becomes energetically preferable. Under these conditions the $\sigma
$-control metaphor implies that artificial restraints (forces) were applied to
the surface charges to enforce uniformity of $\sigma$. \ Elimination of these
artificial restraints would result in transition to an inhomogeneous state. We
will show this and its relation to the sign of the capacitance by considering
an undulating membrane capacitor, an exactly solvable model that contains many
features common to real EDLs, which is thus generally useful for discussing
double layer behavior under critical conditions.

For a membrane capacitor under potential control the onset of instability
results from the steep increase of the electrostatic force $\sim\phi^{2}%
/h^{2}$ where $\phi$ is the applied potential and $h$ is the membrane
thickness. This arises because the system is open: as $h$ decreases, thinning
leads to charge transfer between a battery and the plates of the capacitor
required to maintain the fixed value of the potential drop $\phi\sim\sigma
h=const$. Mathematically, charge transfer is controlled by the term $-q\phi
~$relating the thermodynamic potentials of the open and isolated capacitor
(Eq. \ref{Expand_Isolate}). Thus, both the charge density and the attractive
force increase unrestrainedly as the plates approach one another. At a
critical voltage, no increase of the elastic repulsive force can compensate
for the catastrophic increase of the electrostatic attraction, leading to the
potential-driven instability first noted by Crowley \cite{Cro73}.

In an isolated system this pathway to instability is forbidden. When isolated,
the total plate charge is fixed and uniform system thinning cannot increase
the attractive force, which remains constant, $\sim\sigma^{2}$. Instability
can only result from nonuniform redistribution of the charge density\ in the
plane of the membrane capacitor with an associated inhomogeneity in $h$. Thus
we are led to consider the possible lateral instability of an isolated
flexible membrane capacitor. For $\phi$-control we solved this problem for
electrolyte charge distributions that satisfy the Poisson-Boltzmann
equation\ \cite{ParDorJor98b}. Here we consider a simplified example, a
capacitor in contact with a "perfect conductor," i.e. $\varepsilon
_{solvent}=\infty$, so that the membrane surfaces are isopotentials, a
reliable approximation even for dilute electrolytes \cite{ParDorJor98b}.

Consider the parallel-plate membrane capacitor with its midplane at $\ z=0$
and an unperturbed thickness (when $q=0)\ $equal to $h_{0}$. Charging the
membrane to charge\textbf{\ }density $\sigma$ yields an electrostatic pressure
which gives rise to membrane compression. Introducing the thinning coefficient
$\alpha=h/h_{0}$, with $h$ the thickness of the compressed membrane, the total
energy of the uniform slab is
\begin{equation}
W_{0}=W_{d,0}+W_{e,0}\ \text{where }W_{d,0}=\frac{1~}{2}K_{s}(\alpha
-1)^{2}~\text{and }W_{e,0}=\frac{h}{2\varepsilon\varepsilon_{0}}\sigma^{2};
\label{En_unif}%
\end{equation}
the two terms are the harmonic approximation to the deformation
(stretching-compression) energy and the electrostatic energy respectively,
$K_{s}$\ is the stretching modulus and the index "$0$" refers to a uniformly
deformed membrane.\textbf{\ }The equilibrium membrane thickness $h(\sigma
)=h_{0}\alpha(\sigma)$ is found from the condition
\[
\partial_{\alpha}W=0
\]
leading to the thinning coefficient
\begin{equation}
\alpha(\sigma)=1-\frac{1}{3}(\frac{\sigma}{\sigma_{cr}})^{2}
\label{thinning-charge}%
\end{equation}

\textbf{\ }where%
\begin{equation}
\sigma_{cr}=\sqrt{\frac{2K_{s}\varepsilon\varepsilon_{0}}{3h_{0}}},
\label{dens_crit}%
\end{equation}
which is interpreted in what follows. The transmembrane potential drop and the
corresponding\ inverse differential capacitance are
\begin{equation}
\phi=\frac{1}{\varepsilon\varepsilon_{0}}\sigma h(\sigma)=\frac{1}%
{\varepsilon\varepsilon_{0}}h_{0}\sigma\lbrack1-\frac{1}{3}(\sigma/\sigma
_{cr})^{2}] \label{potential_drop}%
\end{equation}
and%

\begin{equation}
C_{\sigma}^{-1}(\sigma)=\frac{d\phi}{d\sigma}=C_{0}^{-1}[1-~(\sigma
/\sigma_{cr})^{2}], \label{capacitance}%
\end{equation}
where
\[
C_{0}^{-1}=\frac{1}{\varepsilon\varepsilon_{0}}h_{0}%
\]
is the inverse capacitance of a capacitor with the fixed gap $h_{0}$. Eq.
\ref{capacitance} reveals the meaning of $~\sigma_{cr}$. It is the charge
density where the differential capacitance $C_{\sigma}$ becomes infinite;
$C_{\sigma}$ is negative for $|\sigma|\ >\sigma_{cr}$. The corresponding
membrane thickness is
\[
h(\sigma_{cr})=\frac{2}{3}h_{0};
\]
at $\sigma=\sigma_{cr}~$the membrane has thinned by $\sim33\%$ , a value
typical of "relaxing gap" capacitor models
\cite{ParKimFel87,KimKorPar89,ParJor93,ParDorJor96,ParDorJor98b,ParJor2001a}.
The membrane is stable relative to virtual uniform compression (thinning)
under $\sigma$-control (i.e.\ assuming a uniform surface charge density)\ for
all\textit{\ }$\sigma$ including the range where $C<0$. \ This is verified
from Eq. \ref{En_unif}:
\begin{equation}
\partial_{\alpha\alpha}^{2}W|_{\sigma}=\frac{K_{s}}{\alpha}>0. \label{q_stab}%
\end{equation}

We now focus on the energy change, $\Delta W$, for an isolated membrane
capacitor in response to a small charge increment, $\Delta\sigma$, assuming
$\sigma$-control. Using the relation%

\[
\partial_{\sigma\sigma}^{2}W(\sigma)=C^{-1}(\sigma)
\]
we find:%

\begin{equation}
\Delta W=\phi(\sigma)\Delta\sigma+\frac{1}{2}C^{-1}(\sigma)~(\Delta\sigma
)^{2}. \label{energy_var}%
\end{equation}
This equation has important consequences. First virtually separate the
membrane capacitor into equal patches $I$ and $II$, each of area $A/2$ and
permit the charge $\Delta q=A\Delta\sigma/2$ to flow from $I$ \ to $II.\ $For
simplicity neglect boundary effects and assume each charge density is uniform
($\sigma_{1}=\sigma-\Delta\sigma$ and $\sigma_{2}=\sigma+\Delta\sigma$) and
that the patches deform independently, i.e. two membrane capacitors are in
parallel, and\textbf{\ } not elastically coupled. Since the potential is
constant in the plane of the membrane, we find from Eq. \ref{energy_var} that
the total energy change is
\begin{equation}
\Delta W_{12}\ =\frac{A}{2}\ \Delta W_{1}+\frac{A}{2}~\Delta W_{2}~\ =\frac
{A}{2}~C^{-1}~(\Delta\sigma)^{2}. \label{energy_gain}%
\end{equation}
Thus $\Delta W_{12}\,\ $is negative if $C$ $<0$ for the chosen $\sigma$%
.$\ $Put differently, $C<0$ provides a driving force leading to a nonuniform
charge distribution and membrane deformation if the artificial $\sigma
$-control restrictions are relaxed.

The\ energy penalty results from the continuous transition between the
properties of the two membrane\textbf{\ }patches and can be described as a
linear interfacial tension. It is proportional to the length of the border
\ between them and, for large $A$, is negligible relative to $\Delta W_{12}$,
Eq. \ref{energy_gain}. Thus this charge density redistribution and the
corresponding non-uniform deformation of the membrane is possible
energetically for those charge densities that lead to a $C<0$\textbf{\ }domain
\ assuming $\sigma$-control. The appearance of\ \ $C<0$ in treatments that
presume a uniform charge density indicates the system is unstable . This
result is similar to the thermodynamic arguments of Nikitas \cite{Nik91a}
\ who considered equilibrium conditions between two separate surface phases.
\ We will now show that the prediction of negative capacitance under $\sigma
$-control\textbf{\ }also implies that there is the possibility of forming an
inhomogeneous phase under $q$-control.. In our virtual experiment we assumed
the charge density is uniform in each patch. Releasing this restriction
provides other pathways for transition to a non-uniform state. Consider, for
example, membrane stability relative to symmetric undulations, the harmonic
variation of membrane thickness:%

\[
h(x)=h+2~u~\cos(kx),
\]
where $u$ is the amplitude of the undulation of the membrane surfaces; the
corresponding "left" and "right" interfaces are described by the equations
\[
z_{r,l}(x)=\pm~z_{0}(x)\text{ where }~z_{0}(x)=h/2\ +u\ \cos(kx).
\]
\newline This problem has been discussed previously for $\phi$-controlled
systems (see \cite{Cro73,ParDorJor98b,ParJor2001a} and references therein). We
treat a $q$-control environment by fixing the average charge density $q=\sigma
A$ instead of the membrane potential $\phi$.

Some aspects of the solution procedure should be stressed.

(1) Unlike under $\phi$-control, under $q$-control the transmembrane potential
drop $V$\ is not fixed by the external source (battery, potentiostat).
However, the conductive surfaces are still equipotentials and the\ potential
$\phi_{q}$\ is constant on the membrane plane.

(2) The value of $\phi_{q}$\ depends on both the original charge density
$\sigma$\ of the unperturbed membrane and the parameters $u$ and $k$,
characterizing the undulations.

(3) \ $\phi_{q\text{ }}$is then determined as follows:

\qquad(a) The solution\ for a fixed but arbitrary $\phi$\ determines the
potential $v(x,z)$\ within the membrane \cite{ParDorJor98b} (see also
discussion in \cite{And95} and references therein).

\qquad(b) The equation $4\pi\sigma(x)=\varepsilon\nabla_{n}v(x,z_{0}(x))
$\ determines the local charge density $\sigma(x)$, where $\nabla_{n}$\ is the
normal derivative taken at the interface $z_{0}(x)$.

\qquad(c) The total interfacial charge $\widetilde{q}(V)$, is found by
integrating $\sigma(x)$\ over the interface with a weighting factor,
$1/\sqrt{1-[\partial_{x}z_{0}(x)]^{2}}$, that accounts for the membrane
stretching associated with undulations.

\ \ \ \ \ (d) $\phi_{q}$\ is determined from the condition $\widetilde
{q}(V)=\sigma A$:%

\begin{equation}
\phi_{q}=\frac{\sigma}{\varepsilon_{m}\varepsilon_{0}}h(\ 1-~\frac
{k\coth(kh/2)}{h}u^{2}) \label{V_q}%
\end{equation}
with $h=h(\sigma)=h_{0}\alpha(\sigma)$.\ The externally fixed potential $\phi
$\ in the equations for $\phi$-control can now be replaced by $\phi_{q}$,
which completes the solution of the problem for $q$-control.

The membrane's electrostatic energy is then%

\begin{equation}
W_{e}^{q}=W_{e,0}^{q}+W_{e,u}^{q} \label{en_electr_charge}%
\end{equation}
where%

\begin{equation}
W_{e,0}^{q}\mathbf{=}\frac{\sigma^{2}}{\varepsilon_{m}\varepsilon_{0}%
}h\mathbf{(}\sigma\mathbf{)}%
\end{equation}
is the energy of the uniform membrane slab and%

\begin{equation}
W_{e,u}^{q}=-W_{e,0}^{q}\frac{ku^{2}}{h(\sigma)}\coth\mathbf{[}\frac
{kh(\sigma)}{2}\mathbf{]} \label{W_u^q}%
\end{equation}
is the undulatory contribution.\textbf{\ \ }The onset of instability is
determined by competition between the\ decrease of the electrostatic energy,
Eq.\ \ref{W_u^q}, and the corresponding increase\textbf{\ }in membrane
deformation energy averaged in XY-plane, $W_{d,u}$\textbf{. }For simplicity,
we consider small\textbf{\ }$k$\textbf{\ }(the long- wavelength limit)
$kh<<1.$\textbf{\ }Similar to \cite{ParDorJor98b} where a slightly different
form of $W_{d,0},$\ Eq. \ref{En_unif}, was used, we can represent $W_{d,u}%
$\bigskip\ as%

\begin{equation}
W_{d,u}\sim\frac{K_{s}u^{2}}{h_{0}^{2}}[1+O((kh)^{2})]. \label{W_du}%
\end{equation}
Higher order terms in $(kh)^{2}$\textbf{\ }arise from surface tension and
bending contributions to the elastic energy and are neglected. They are
analogs to the non-uniform interfacial contributions of the previous example.
In the same limit Eq. \ref{W_u^q} can be represented as\textbf{\ }%
\begin{equation}
W_{e,u}^{q}\sim-2W_{e,0}^{q}u^{2}[\frac{1}{h(\sigma)^{2}}+O((kh)^{2})].
\label{W_eu}%
\end{equation}
The uniform distribution becomes unstable when
\[
W_{u}=W_{d,u}+W_{e,u}^{q}\leq0.
\]
Substituting Eqs. \ref{W_u^q} and \ref{thinning-charge} we find the condition
of instability:%

\[
\mid\overline{\sigma}\mathbf{|\ \geq\sigma}_{cr},\text{ }%
\]
which is equivalent to\textbf{\ }%
\[
C_{q}^{-1}\leq0.
\]
Our thought experiment presumed that the conditions for $\sigma$-control ,
i.e. uniformity of $\sigma$, could\textbf{\ }be arbitrarily relaxed anywhere
within the negative $C$\ domain. In reality there is no way to enforce
uniformity when\ the system is unstable; thus the distribution\ spontaneously
becomes inhomogeneous at the edge of this domain with critical point
\ $\sigma=\sigma_{cr}$, where%
\begin{equation}
C_{q}^{-1}(\sigma_{cr})=0; \label{1/C=0}%
\end{equation}
the transition actually occurs a bit earlier, at a point roughly determined by
a Maxwell construction \cite{ParJor93}.

We have analyzed a simplified model corresponding to a concentrated
electrolyte (Debye length $\lambda_{D}\rightarrow0)$. \ Further analysis based
on our previous work shows that Eq. \ref{1/C=0} also determines the onset of
instability for finite $\lambda_{D}$ \cite{ParDorJor98b}. This result is
valuable on its own. As already indicated the Poisson-Boltzmann approximation
and other local statistical models do not predict NC\textbf{\ }(see
\cite{FelParVor86b,ParJor93} and references therein). Eqs. \ref{cap_relaxed}
and \ref{C<0 _cond}\ show that for these models the rate of gap contraction
with charging, $d^{\prime}(\sigma),$ is always less than $1/|\sigma|$ and thus
$C_{\sigma}$ is always positive. Consequently, this class of models would not
satisfy the criteria suggested in \cite{GonJimMes2004} linking model quality
to the appearance of a $C_{\sigma}<0$ domain.\ Our results indicate
that\ adding another relaxation mechanism immediately leads to the appearance
of a $C_{\sigma}<0$\ domain and instability. This and our earlier discussion
of the electronic models illustrates that anomalies are more typical than
expected based on purely ionic models\ with immobile\ charged
plates.\textbf{\ }

In previous analysis \cite{ParDorJor96} we considered two elastically coupled
membrane capacitors, with the extra term in the deformational energy
$\sim\alpha\ (u_{1}-u_{2})^{2}$\textbf{\ }accounting\textbf{\ }for the
non-uniformity penalty (differential thinning of the patches). Depending on
the coupling constant $\alpha$,\ this system could exhibit a $C<0$%
\textbf{\ }domain before transition to a nonuniform state. Our present
discussion implies such a picture is unrealistic. In terms of the first model,
the constant $\alpha$\ must be proportional to the width of the transition
region relative to the\textbf{\ }area of the patches\textbf{\ }$A$%
\textbf{\ }and thus can become infinitesimal if $A$ is sufficiently large. In
addition, as shown in the second example, the non-uniformity contributions
become insignificant at small $k$\ which provides a reasonable
pathway\textbf{\ }for the onset of instability.

Finally, we reiterate the major difference between $q$-\ and $\phi$-control
for the onset of instability. Under $\phi$-control (an open system connected
to a potentiostat) stability is lost simultaneously for both uniform
deformation and undulations \cite{Cro73,ParDorJor98b}. In contrast, a
$q$-controlled (isolated) system is always stable with respect to uniform
deformation (see Eq. \ref{q_stab} and the corresponding discussion) and only
loses stability in transiting to a nonuniform state. An important consequence
is that in a $\phi$-driven transition the original and final phases correspond
to different values of $q$\ while under $q$-control only the local charge
density can change provided its average value is fixed.

\section{Perspectives for further study}

It is by now well established that the capacitance can be negative for
uniformly charged surfaces under the artificial conditions of $\sigma
$-control. Many statistical ionic models have demonstrated this anomaly (see
\cite{Tor92,BodHenPla2004,GonJimMes2004} and references therein). While in the
1980s many would have viewed such predictions as model faults, the most recent
view \cite{GonJimMes2004,BodHenPla2004} represents a dramatic change of mind,
and it is even suggested that the prediction of $C<0$ must be considered as a
criterion validating an ionic model \cite{GonJimMes2004} rather than an
imperfection. We do not disagree. \ In fact, it accords with our observation
\cite{FelParVor86b,KimKorPar89,ParJor93,ParDorJor96} that such predictions are
typical of models jointly accounting for various contributions (electronic,
ionic, etc.) to the charging induced relaxation of the effective gap of the
interfacial capacitor. \ 

What is the physical significance of such predictions? Our analysis indicates
that they imply an instability with respect to a transition to an
inhomogeneous state. To analyze the consequences, the artificial assumption of
a uniform surface charge distribution must be dropped, given that in a real
metallic electrodes the electrons are free to move and thus surface charge
density can become, at least in principle, laterally non-uniform. This
non-uniformity can be even more pronounced in soft media such as lipid
bilayers. The value of the models predicting this anomaly, emphasized in
\cite{GonJimMes2004}, is even greater since they are candidates for analyzing
such instability.

We must re-emphasize that in our usage the terms "instability" and
"transition" are not related to real interfacial critical phenomena. While the
phase transition actually occurs before $C_{\sigma}$\ becomes negative, we
permit the system to enter this domain by artificially maintaining $\sigma
$-control. Relaxing the uniformity constraint at any $\sigma$\ within the
$C<0$\textbf{\ }domain leads immediately to a transition to an inhomogeneous
state of fixed $q$, $q=\sigma A.$\ Although the initial ($\sigma$-controlled)
state is artificial, the final stable inhomogeneous phase (if it exists) is
real since the equilibrium state is unique.\ Thus, our approach is useful for
testing and developing statistical models that describe both uniform
($C>0)~$and inhomogeneous (regions with $C\ <0$\ under $\sigma$-control) \ phases.

Predicting instability does not necessarily imply that the model describes the
formation of a new stable inhomogeneous phase. As the transition can be
accompanied by a substantial local increase of the charge density and a
corresponding local increase of the ionic density in the EDL, a model must
have a stabilization mechanism that interrupts the\textbf{\ }propagation of
instability. This would permit formation of new stable phase before the
condition of "ideal polarizability" is broken and interfacial charge transfer
occurs. Ionic size and correlation effects in the electrolyte must be
important here.

Although the condition $C<0$, obtained for a primitive ionic model of the EDL,
can be an important factor leading to instability, it is not a necessary
condition. Even with ionic models that by themselves do not lead to
instability (such as GCS model) the addition of other mechanisms of
relaxation, such as a displacement of the "electronic plate" of the
interfacial capacitor, can lead to $C_{\sigma}<0$ and thus trigger the
instability \cite{FelParVor86b,ParDorJor96}. In other words, this anomaly
should be even more commonplace than is implied by ionic model studies.

Another important question is a comparison of phase transitions predicted for
the open and the isolated systems. While in the first case the transition can
be accompanied by charging the electrodes, in the second case the lateral
variation of charge keeps the total charge fixed. Finally, the inherent
inhomogeneity (roughness) of an electrode surface (especially with respect to
solid electrodes) must be considered. The influence of the roughness on the
equilibrium properties of EDL is well established \cite{DaiKorUrb96}, and its
possible effect on the surface phase transition can also be a promising field
for further research.

\section*{Acknowledgement}

Work supported by a grant from the National Institutes of Health, GM28643.
M.B.P. wishes to thank Dr. V.J. Feldman for many fruitful discussions and
Professor A.M. Brodsky for suggesting the possibility of charge density-type
phases in double layers (private communication, $\sim$ 1988).

\bigskip

\bibliographystyle{plain}

\end{document}